\documentclass[aps,preprint]{revtex4}%
\usepackage{amsfonts}
\usepackage{amsmath}
\usepackage{amssymb}
\usepackage{graphicx}
\usepackage[usenames]{color}%
\providecommand{\U}[1]{\protect\rule{.1in}{.1in}}

\begin{document}
\preprint{ }
\title{Diversified vortex phase diagram for a rotating trapped two-band Fermi gas in
the BCS-BEC crossover}
\author{S. N. Klimin}
\altaffiliation{Also at: Department of Theoretical Physics, State University of Moldova}

\affiliation{TQC, Universiteit Antwerpen, Universiteitsplein 1, B-2610 Antwerpen, Belgium}
\author{J. Tempere}
\altaffiliation{Also at: Lyman Laboratory of Physics, Harvard University}

\affiliation{TQC, Universiteit Antwerpen, Universiteitsplein 1, B-2610 Antwerpen, Belgium}
\author{M. V. Milo\v{s}evi\'{c}}
\thanks{Email: milorad.milosevic@uantwerpen.be}
\affiliation{Departement Fysica, Universiteit Antwerpen, Groenenborgerlaan 171, B-2020
Antwerpen, Belgium}

\begin{abstract}
We report the equilibrium vortex phase diagram of a rotating two-band Fermi
gas confined to a cylindrically symmetric parabolic trapping potential, using
the recently developed finite-temperature effective field theory [Phys. Rev. A
\textbf{94}, 023620 (2016)]. A non-monotonic resonant dependence of the free
energy as a function of the temperature and the rotation frequency is revealed
for a two-band superfluid. We particularly focus on novel features that appear
as a result of interband interactions and can be experimentally resolved. The
resonant dependence of the free energy is directly manifested in vortex phase
diagrams, where areas of stability for both integer and fractional vortex
states are found. The study embraces the BCS-BEC crossover regime and the
entire temperature range below the critical temperature $T_{c}$. Significantly
different behavior of vortex matter as a function of the interband coupling is
revealed in the BCS and BEC regimes.

\end{abstract}
\date{\today}
\maketitle

\section{Introduction \label{sec:intro}}

Quantum gases constitute a remarkable testing ground for the theory of the
superfluid state and its various macroscopic excitations, such as vortices and
solitons. Vortices are stabilized in an atomic gas by rotating the gas, just
as vortices in superconductors can be stabilized by a magnetic field: the
Coriolis force acts on a particle in a rotating frame of reference in the same
way as the Lorentz force acts on a charged particle in a magnetic field
\cite{Bloch2008}. Consequently, vortices and many-vortex states in rotating
trapped superfluid Fermi gases have become a subject of an intense
experimental \cite{zw1,zw2,zw3,zw4} and theoretical
\cite{Clancy,Riedl,Bruun,Cozzini,Urban,Warringa,Warringa2,Wei2012,Simonucci2015,RotVort1}
research during last two decades.

Since recently, this line of research has become of particular interest in
multicomponent quantum systems. For example, multiband superconductors, such
as MgB$_{2}$, were intensely studied both experimentally and theoretically
over the last decade. However, interaction parameters can hardly be tuned in
solid-state systems. Contrary to superconductors, interactions in ultracold
Fermi gases can be controlled, and a broad range of regimes from BCS to BEC
can be realized. This stimulated a theoretical interest to multiband quantum
gases in anticipation of future experiments \cite{Iskin2006,Iskin2005,He2015}.
Recently, a two-band Fermi superfluid has been successfully created in an
ultracold gas of $^{173}$Yb atoms \cite{Pagano2015,Hofer2015} using the
orbital Feshbach resonance predicted in Ref. \cite{Zhang2015} (see also Ref.
\cite{Xu2016}). This makes a theoretical investigation of different phenomena
in multiband Fermi gases (vortices, solitons, etc.) timely and important.
Vortices in multiband superconducting and superfluid systems are particularly
interesting due to a rich variety of observable phenomena, such as fractional
vortex states that occur when winding numbers in different band-components of
the condensate are not equal. Fractional vortices in multiband superconductors
have been investigated fairly thoroughly \cite{M2012,Chibotaru,Babaev2002}.
Concerning the BCS-BEC crossover, there was a widely spread opinion during a
long time that superconductors cannot be realized away from the BCS regime.
However recently the BCS-BEC crossover has been successfully reached in
superconducting FeSe \cite{Kasahara}. Moreover, vortex matter in multiband
superconductors in the BCS-BEC crossover regime has also attracted much
attention \cite{Sun}. Vortex states in multiband quantum atomic gases have
been studied to a far lesser extent.

Therefore, in this work, the subject of our interest are fractional vortices
in two-band Fermi gases of ultracold atoms in the BCS-BEC crossover. Although
there is some similarity between superconductors and condensed atomic Fermi
gases, the analogy is not complete. For one difference, cold gases certainly
require an independent treatment by specific methods suitable in the entire
BCS-BEC crossover range.

Recently, the stability of different vortex states in a rotating trapped
one-band Fermi gas has been theoretically studied in Refs.
\cite{Simonucci2015,RotVort1} using, respectively, the coarse-grained
Bogoliubov -- de Gennes (BdG) theory \cite{Simonucci2014} (first applied to
atomic Fermi gases in \cite{Sensarma}) and the recently developed
finite-temperature effective field theory (EFT) \cite{KTD,KTLD2015}. The
finite-temperature EFT results agree with the results of the BdG theory and
experiment for different manifestations: collective excitations and vortices
\cite{RotVort1,KTLD2015}, and solitons \cite{KTD2014,LAKT2015}. The finite
temperature EFT is aimed to find analytic results whenever possible. For
example, for dark solitons in condensed Fermi gases, the finite-temperature
EFT provides exact analytic solutions of the soliton equation of motion
\cite{KTD2014}, while the Bogoliubov -- de Gennes equations for the same
problem have been solved only numerically.

Besides our studies, there are several modifications of the effective field
theory of condensed Fermi gases described in different publications and
related to different ranges of external parameters (e.~g., temperature and
scattering length). They are developed either for the close vicinity to the
critical temperature \cite{deMelo1993} or for the case $T=0$ (e.~g.,
\cite{Diener2008,Marini1998,Schakel}). As analyzed in Ref. \cite{KTLD2015},
the present finite-temperature EFT agrees with preceding works in all these
limiting cases.

At present, experimental data on vortices in two-band superfluid atomic Fermi
gases are still lacking, in spite of the expected new physics stemming from
the interband interactions in such a system. We report the first such
theoretical study, to pave the way for future experiments with two- and
multiband atomic Fermi superfluids. This work builds on the research performed
in Ref. \cite{RotVort1}, with an extension to two-band fermionic systems. The
main goal of the present investigation is to reveal novel vortex phenomena
which can appear in a two-band Fermi gas, that are arguably easy to verify
experimentally. More specifically, we study the evolution of equilibrium
vortex states when varying the temperature and the interband coupling
strength, as well as the frequency of rotation, to identify regions of
stability for fractional vortices, clusters of non-composite vortices, and
multivortex states. Two variants are considered: (1) the \textquotedblleft
canonical ensemble\textquotedblright\ case when numbers of particles in each
band are fixed separately, and (2) the \textquotedblleft grand canonical
ensemble\textquotedblright\ case, when the numbers of particles are determined
from the common chemical potential.

The paper is organized as follows. In Sec. \ref{sec:EFT}, the used method and
approximations are described. In Sec. \ref{sec:Results}, the parameters of
state and the vortex phase diagrams are analyzed in detail. Our results are
summarized in Sec. \ref{sec:Conclusions}.

\section{Method \label{sec:EFT}}

We apply the finite-temperature EFT developed in Ref. \cite{KTLD2015} to
vortices and many-vortex states in a two-band Fermi gas with $s$-wave pairing.
The fermion system is confined to a cylindrically symmetric parabolic trapping
potential with confinement frequency $\omega_{0}$. This setup can be relevant
for ultracold fermionic atoms in elongated traps, as, e. g., in Refs.
\cite{zw1,zw2,zw3,zw4}. The stabilization of vortices is achieved by rotating
the fermionic system with an angular frequency $\omega$. The rotation is
incorporated in the EFT in the same manner as in Ref. \cite{RotVort1}.

The details of the finite-temperature EFT with a discussion of the
approximations used and the range of applicability of the method are readily
available in Ref. \cite{KTLD2015}. The incorporation of rotation in the EFT
and the extension of the formalism to a two-band rotating Fermi gas have been
already described in Ref. \cite{RotVort1}. Consequently, in this section we
reproduce the formalism only briefly, because detailed derivations,
discussions and proofs can be found in Refs. \cite{RotVort1,KTLD2015}.

The treatment of rotating trapped Fermi gases is performed within the
path-integral formalism in the space of anticommuting fermion fields, see, e.
g., Refs. \cite{deMelo1993,Diener2008}. Throughout the treatment, we use units
such that $\hbar=1$, the fermion mass for a \textquotedblleft
strong\textquotedblright\ band $m_{1}=1/2$, and the Fermi energy for a
non-interacting fermion system $E_{F}=1$. Note that there is a difference
between the Fermi energy for a Fermi gas in bulk and in a trapping potential.
In the present system of units, $E_{F}$ is the global rather than a local
parameter, i.~e., $E_{F}$ is not coordinate-dependent. For example, the Fermi
energy for a gas trapped to a 3D parabolic confinement potential with
confinement frequencies $\omega_{x},\omega_{y},\omega_{z}$ is $E_{F}%
\approx\hbar\left(  3\omega_{x}\omega_{y}\omega_{z}N\right)  ^{1/3}$ when $N$
is sufficiently large. The parameter having the dimensionality of the Fermi
wave vector is formally determined as $k_{F}\equiv\sqrt{2mE_{F}}/\hbar$.

The thermodynamic quantities are calculated on the basis of the partition
function,
\begin{equation}
\mathcal{Z}\propto\int\mathcal{D}\left[  \bar{\psi},\psi\right]  e^{-S}.
\end{equation}
Here, $S$ is the action functional for a two-band fermionic system
\cite{KTLD2015}:%
\begin{equation}
S=\sum_{j=1,2}S_{0,j}+\int_{0}^{\beta}d\tau\int d\mathbf{r}~U\left(
\mathbf{r},\tau\right)  , \label{S1}%
\end{equation}
where $\beta=1/\left(  k_{B}T\right)  $, $T$ is the temperature, $k_{B}$ is
the Boltzmann constant, and $S_{0,j}$ is the free-fermion action for the
$j$-th band,%
\begin{equation}
S_{0,j}=\int_{0}^{\beta}d\tau\int d\mathbf{r}\sum_{\sigma=\uparrow,\downarrow
}\bar{\psi}_{\sigma,j}\left(  \frac{\partial}{\partial\tau}+H_{\sigma
,j}\right)  \psi_{\sigma,j}, \label{S01}%
\end{equation}
with the one-particle Hamiltonian in the rotating frame of reference, which
allows for independent populations in different bands and different spins
$\sigma$, given by%
\begin{equation}
H_{\sigma,j}=-\frac{\left(  \nabla-i\mathbf{A}_{j}\left(  \mathbf{r}\right)
\right)  ^{2}}{2m_{j}}+\frac{m_{j}\left(  \omega_{0}^{2}-\omega^{2}\right)
}{2}\left(  x^{2}+y^{2}\right)  -\mu_{\sigma,j}, \label{Hj}%
\end{equation}
where $\mathbf{A}_{j}\left(  \mathbf{r}\right)  $ is the rotational\ vector
potential $\mathbf{A}_{j}\left(  \mathbf{r}\right)  =m_{j}\left[
\boldsymbol{\omega}\times\mathbf{r}\right]  $, and $\boldsymbol{\omega}%
\equiv\omega\mathbf{e}_{z}$ is the rotation vector. The fermion-fermion
interaction $U\left(  \mathbf{r},\tau\right)  $ for a two-band system assumes
both intraband and interband contact interactions,%
\begin{align}
U  &  =%
{\textstyle\sum\nolimits_{j=1,2}}
g_{jj}\bar{\psi}_{\uparrow,j}\bar{\psi}_{\downarrow,j}\psi_{\downarrow,j}%
\psi_{\uparrow,j}\nonumber\\
&  +g_{12}^{\left(  a\right)  }\left(  \bar{\psi}_{\uparrow,1}\psi
_{\uparrow,1}\bar{\psi}_{\downarrow,2}\psi_{\downarrow,2}+\bar{\psi
}_{\downarrow,1}\psi_{\downarrow,1}\bar{\psi}_{\uparrow,2}\psi_{\uparrow
,2}\right) \nonumber\\
&  +g_{12}^{\left(  p\right)  }\left(  \bar{\psi}_{\uparrow,1}\psi
_{\uparrow,1}\bar{\psi}_{\uparrow,2}\psi_{\uparrow,2}+\bar{\psi}%
_{\downarrow,1}\psi_{\downarrow,1}\bar{\psi}_{\downarrow,2}\psi_{\downarrow
,2}\right)  , \label{U}%
\end{align}
with intraband coupling constants $g_{jj}$($j=1,2$) and interband coupling
constants $g_{12}^{\left(  a\right)  }$ and $g_{12}^{\left(  p\right)  }$
which describe interband contact interactions of density fluctuations with
antiparallel and parallel spins, respectively.

The finite-temperature EFT action for a one- and two-band system of
interacting fermions with $s$-wave pairing has been formulated in Ref.
\cite{KTLD2015}. Here, details of the derivation of the effective field action
are given in Appendix A. After the Hubbard-Stratonovich transformation using
bosonic pair fields $\Psi_{1},\Psi_{2}$, the integration over fermion fields
and the gradient expansion (assuming slow variation of the pair field in time
and space), the resulting action functional for a two-band system
$S_{EFT}^{\left(  2b\right)  }$ takes the form:%
\begin{equation}
S_{EFT}^{\left(  2b\right)  }=\sum_{j=1,2}S_{EFT}^{\left(  j\right)  }%
-\int_{0}^{\beta}d\tau\int d\mathbf{r~}\frac{\sqrt{m_{1}m_{2}}}{4\pi}%
\gamma\left(  \bar{\Psi}_{1}\Psi_{2}+\bar{\Psi}_{2}\Psi_{1}\right)  ,
\label{Seff2B}%
\end{equation}
where $\gamma$ is the strength of the interband coupling expressed through the
interband scattering lengths,%
\begin{equation}
\gamma=2\left(  \frac{1}{a_{s,12}^{\left(  a\right)  }}-\frac{1}%
{a_{s,12}^{\left(  p\right)  }}\right)  . \label{gam}%
\end{equation}
The one-band effective field action $S_{EFT}^{\left(  j\right)  }$ for the
$j$-th band is determined by:%
\begin{align}
S_{EFT}^{\left(  j\right)  }  &  =\int_{0}^{\beta}d\tau\int d\mathbf{r}\left[
\Omega_{s,j}\left(  \left\vert \Psi_{j}\right\vert ^{2}\right)  +\frac{D_{j}%
}{2}\left(  \bar{\Psi}_{j}\frac{\partial\Psi_{j}}{\partial\tau}-\frac
{\partial\bar{\Psi}_{j}}{\partial\tau}\Psi_{j}\right)  \right. \nonumber\\
&  +Q_{j}\frac{\partial\bar{\Psi}_{j}}{\partial\tau}\frac{\partial\Psi_{j}%
}{\partial\tau}-\frac{R_{j}}{2\left\vert \Psi_{j}\right\vert ^{2}}\left(
\frac{\partial\left(  \left\vert \Psi_{j}\right\vert ^{2}\right)  }%
{\partial\tau}\right)  ^{2}+C_{j}\left(  \nabla_{\mathbf{r}}\bar{\Psi}%
_{j}\cdot\nabla_{\mathbf{r}}\Psi_{j}\right)  -E_{j}\left(  \nabla_{\mathbf{r}%
}\left(  \left\vert \Psi_{j}\right\vert ^{2}\right)  \right)  ^{2}\nonumber\\
&  \left.  +iD_{j}\mathbf{A}_{j}\cdot\left(  \bar{\Psi}_{j}\nabla_{\mathbf{r}%
}\Psi_{j}-\Psi_{j}\nabla_{\mathbf{r}}\bar{\Psi}_{j}\right)  \right]  .
\label{Seff}%
\end{align}
The coordinate-dependent thermodynamic potential $\Omega_{s,j}\left(
\left\vert \Psi_{j}\right\vert ^{2}\right)  $ and the coefficients of the
gradient expansion given in Refs. \cite{RotVort1,KTLD2015} and explained in
detail in Appendix A. The rotation is incorporated in the effective action as
described in Ref. \cite{RotVort1}. It leads to the appearance of the linear
gradient term $iD_{j}\mathbf{A}_{j}\cdot\left(  \bar{\Psi}_{j}\nabla
_{\mathbf{r}}\Psi_{j}-\Psi_{j}\nabla_{\mathbf{r}}\bar{\Psi}_{j}\right)  $ in
the effective bosonic action, which vanishes in the absence of rotation. As
shown in Ref. \cite{RotVort1}, the obtained extension of the effective bosonic
action to a rotating fermion system is in agreement with results of the
functional renormalization group theory \cite{Boettcher,Diehl}.

\section{Confined vortex states in a two-band Fermi gas \label{sec:Results}}

\subsection{Conditions for vortex stability}

One of the most interesting consequences of rotation in a two-band Fermi
system is the formation of fractional vortices. The fractional vortices can be
energetically favorable only at rather small interband coupling strength,
because the Josephson coupling energy [the last term in (\ref{Seff2B})]
penalizes phase differences between the condensates in different bands.
However, two-band Fermi gases with vanishingly small (and even zero) interband
coupling can be prepared in practice -- for example, when the two condensates
are spatially separated. Therefore in the present work we study vortices in
two-band Fermi gases with a particular attention to the range of small
Josephson coupling strengths.

The stability of different vortex states is analyzed using the free energy
$F^{\left(  2b\right)  }$, which is obtained from the effective action
(\ref{Seff}) assuming pair fields to be stationary (time-independent),%
\begin{equation}
F^{\left(  2b\right)  }=\sum_{j=1,2}F^{\left(  j\right)  }-\int d\mathbf{r~}%
\frac{\sqrt{m_{1}m_{2}}}{4\pi}\gamma\left(  \bar{\Psi}_{1}\Psi_{2}+\bar{\Psi
}_{2}\Psi_{1}\right)  , \label{F2b}%
\end{equation}
where $F^{\left(  j\right)  }$ is the contribution to the free energy from the
$j$-th band-component of the condensate:%
\begin{align}
F^{\left(  j\right)  }  &  =\int d\mathbf{r}\left[  \Omega_{s,j}\left(
\left\vert \Psi_{j}\right\vert ^{2}\right)  +C_{j}\left(  \nabla_{\mathbf{r}%
}\bar{\Psi}_{j}\cdot\nabla_{\mathbf{r}}\Psi_{j}\right)  \right. \nonumber\\
&  \left.  -E_{j}\left(  \nabla_{\mathbf{r}}\left(  \left\vert \Psi
_{j}\right\vert ^{2}\right)  \right)  ^{2}+iD_{j}\mathbf{A}_{j}\cdot\left(
\bar{\Psi}_{j}\nabla_{\mathbf{r}}\Psi_{j}-\Psi_{j}\nabla_{\mathbf{r}}\bar
{\Psi}_{j}\right)  \right]  . \label{Fj}%
\end{align}

The pair field $\Psi_{j}\left(  \mathbf{r}\right)  $ for a particular vortex
state with $n$ vortices in the $j$-th band-component of the condensate is
described here by the variational function, which is a product of the
background pair field $\Delta_{j}\left(  r\right)  $ and the vortex factor:%
\begin{equation}
\Psi_{j}\left(  \mathbf{r}\right)  \equiv\Psi_{j}\left(  \mathbf{r},\left\{
a_{\nu,j},\theta_{\nu,j}\right\}  \right)  =\Delta_{j}\left(  r\right)
\sum_{\nu=1}^{n}a_{\nu,j}\left(  \mathbf{r}\right)  e^{i\theta_{\nu,j}\left(
\mathbf{r}\right)  }, \label{vort}%
\end{equation}
where $a_{\nu,j}\left(  \mathbf{r}\right)  $ and $\theta_{\nu,j}\left(
\mathbf{r}\right)  $ are, respectively, the relative amplitude and the phase
for the $\nu$-th vortex in the $j$-th band component of the condensate. The
phase for a vortex $\theta_{\nu,j}\left(  \mathbf{r}\right)  $ coincides with
the polar angle calculated in the frame of reference related to the vortex
center. The vortex amplitudes in the present work, as well as in Ref.
\cite{RotVort1}, are modeled by variational functions,%
\begin{equation}
a_{\nu,j}\left(  \mathbf{r}\right)  =\tanh\left(  \frac{\left\vert
\mathbf{r-r}_{\nu,j}\right\vert }{\sqrt{2}\xi_{j}}\right)  . \label{var1}%
\end{equation}
The parameter $\xi_{j}$ has the sense of the vortex healing length. The
healing lengths and the positions of vortex centers $\mathbf{r}_{\nu,j}$ are
variational parameters for vortex states. Their optimal values are found by
the minimization of the free energy (\ref{F2b}) after substituting the
variational function (\ref{vort}) there. The conditions of stability for
vortex states are determined here, as in Ref. \cite{RotVort1}, by considering
the difference between the free energies with and without vortices:%
\begin{equation}
\delta F\equiv F\left[  \left\{  \Psi_{j}\left(  \mathbf{r}\right)  \right\}
\right]  -F\left[  \left\{  \Delta_{j}\left(  r\right)  \right\}  \right]  .
\label{dF}%
\end{equation}
The parametric ranges of existence for vortex configurations are determined by
comparison of free energies corresponding to different vortex states,
including also the state without vortices, where $\Psi_{j}\left(
\mathbf{r}\right)  =\Delta_{j}\left(  r\right)  $, and choosing a vortex state
with the lowest free energy.

The background distribution of the pair field $\Delta_{j}\left(  r\right)  $
without vortices is determined using normalization conditions for the fermion
density. We consider here two possible constraints of the numbers of fermions
in two bands. Under the first condition, total numbers of particles in each
band are fixed. The alternative condition is the \textquotedblleft grand
canonical\textquotedblright\ setting, when the populations in the two bands
have been relaxed to values determined through equal chemical potentials in
the two band-components of the condensate. At present, the ensemble with fixed
numbers of particles seems to be more easily achieved in experiments with
atomic Fermi gases than the grand canonical ensemble with equal chemical
potentials. However, it is interesting to consider them both, in order to
understand the influence of these constraints on vortex states.

In order to make clear the difference between the two aforesaid ensembles, we
describe the normalization conditions for trapped two-band Fermi condensates
in detail. Since we consider systems where the scale of the confinement
potential is sufficiently large with respect to the vortex size, the
normalization conditions are applied within the local density approximation
through coordinate-dependent thermodynamic parameters. For a one-band Fermi
gas trapped in a cylindrically symmetric confinement potential, the
normalization condition in the local density approximation reads,
\begin{equation}
2\pi\int_{0}^{\infty}n\left(  \beta,\mu\left(  r\right)  ,\Delta\left(
r\right)  \right)  rdr=N, \label{norm1B}%
\end{equation}
where $n\left(  \beta,\mu\left(  r\right)  ,\Delta\left(  r\right)  \right)  $
is the fermion density depending on the radial variable $r=\sqrt{x^{2}+y^{2}}$
through the \emph{background} pair field $\Delta\left(  r\right)  $ (we
neglect here the feedback of vortices to the density normalization) and the
chemical potential, according to (\ref{Hj}),%
\[
\mu\left(  r\right)  =\mu\left(  0\right)  -\frac{m}{2}\left(  \omega_{0}%
^{2}-\omega^{2}\right)  r^{2}.
\]
The coordinate-dependent gap parameter is determined through the local gap
equation \cite{KTLD2015},
\begin{equation}
\int\frac{d\mathbf{k}}{\left(  2\pi\right)  ^{3}}\left(  \frac{\sinh\beta
E_{\mathbf{k}}}{2E_{\mathbf{k}}\left(  \cosh\beta E_{\mathbf{k}}+\cosh
\beta\zeta\right)  }-\frac{m}{k^{2}}\right)  +\frac{m}{4\pi a_{s}}=0,
\label{gapeq1}%
\end{equation}
consistently with the normalization condition (\ref{norm1B}). Here,
$E_{\mathbf{k}}=\sqrt{\left(  \frac{k^{2}}{2m}-\mu\left(  r\right)  \right)
^{2}+\Delta\left(  r\right)  ^{2}}$ is the Bogoliubov excitation energy,
$\mu\left(  r\right)  =\left(  \mu_{\uparrow}+\mu_{\downarrow}\right)  /2$ and
$\zeta=\left(  \mu_{\uparrow}-\mu_{\downarrow}\right)  /2$ are, respectively,
the averaged chemical potential and the difference of chemical potentials for
the \textquotedblleft spin up\textquotedblright\ and \textquotedblleft spin
down\textquotedblright\ species.

The fermion density $n$ entering normalization condition (\ref{norm1B}) can be
calculated in different approximations, e. g., the mean-field or Gaussian pair
fluctuation approximation, as in Ref. \cite{OurPRA2008}. In the present work,
we restrict the normalization of the density by the mean-field approximation,
because account of fluctuations in the background parameters can lead only to
a rescaling of phase diagrams, without changing their shape and sequences of
areas of stability for vortex states. It should be noted that the gap equation
has the same form (\ref{gapeq1}) with and without account of fluctuations
\cite{Diener2008,OurPRA2008}.

For a two-band system, coordinate-dependent parameters of state are determined
from two normalization conditions,%
\begin{equation}
2\pi\int_{0}^{\infty}n_{j}\left(  \beta,\mu_{j}\left(  r\right)  ,\Delta
_{j}\left(  r\right)  \right)  rdr=N_{j}\quad\left(  j=1,2\right)
\label{norm2B}%
\end{equation}
together with the coupled set of two gap equations for the two-band Fermi gas
given by Eq. (28) of Ref. \cite{KTLD2015},
\begin{align}
&  \left[  \int\frac{d\mathbf{k}}{\left(  2\pi\right)  ^{3}}\left(  \frac
{1}{2E_{\mathbf{k},j}}\frac{\sinh(\beta E_{\mathbf{k},j})}{\cosh(\beta
E_{\mathbf{k},j})+\cosh(\beta\zeta_{j})}-\frac{m_{j}}{k^{2}}\right)
+\frac{m_{j}}{4\pi a_{s,j}}\right]  \Delta_{j}\nonumber\\
&  +\frac{\sqrt{m_{1}m_{2}}}{4\pi}\gamma\Delta_{3-j}=0. \label{gapset}%
\end{align}

The difference between two aforesaid ensembles consists in the following.
Under the first condition, numbers of particles $N_{j}$ in (\ref{norm2B}) are
fixed. The chemical potentials $\mu_{j}\left(  r\right)  $ are determined from
these two equations, taking into account (\ref{gapset}) which in fact depends
on both $\mu_{1}$ and $\mu_{2}$. In the \textquotedblleft grand
canonical\textquotedblright\ setting, $N_{1}$ and $N_{2}$ are not fixed
separately. Instead, we apply (\ref{gapset}) with the normalization condition
for the total number of particles $N\equiv N_{1}+N_{2}$,%
\begin{equation}
2\pi\sum_{j=1,2}\int_{0}^{\infty}n_{j}\left(  \beta,\mu\left(  r\right)
,\Delta_{j}\left(  r\right)  \right)  rdr=N, \label{Norm2B-1}%
\end{equation}
where $\mu\left(  r\right)  $ is the same for both band-components of the
condensate. Consequences of this difference, in particular the depletion of
the \textquotedblleft weak\textquotedblright\ band at low temperatures for the
\textquotedblleft grand canonical\textquotedblright\ setting, were considered
in Ref. \cite{KTLD2015}.

The free energy and the normalization integrals are calculated including both
the superfluid core and the normal phase which are spatially separated. It is
assumed here that they both rotate with the frequency imposed by a stirring
field. A treatment beyond this approximation may lead only to a marginal
correction to the obtained results without changing the picture, because we
consider vortices near the trap center, i. e., far from the boundary between
the superfluid and normal phases. The same assumption has been substantiated
and applied in Refs. \cite{Simonucci2015,RotVort1}.

The term \textquotedblleft fractional vortex\textquotedblright\ means that
winding numbers for vortex states in two bands are not the same. In order to
classify different vortex configurations, we label the \textquotedblleft
strong\textquotedblright\ band (with a larger inverse scattering length
$1/a_{s}$) by index 1, and the \textquotedblleft weak\textquotedblright\ band
by index 2. Following Ref. \cite{M2012} we use the notation $\left(
n_{1},n_{2},n_{c}\right)  $ for the classification of vortex states, where
$n_{1}$ and $n_{2}$ are, respectively, the winding numbers in the first and
second band-components of the condensate, and $n_{c}$is the number of
composite vortices, i.~e., integer vortices pinned to each other in two bands.
Correspondingly, the clouds in two band-components of the condensate
classified by the set of indices $\left(  n_{1},n_{2},n_{c}\right)  $ look as
or multi-vortex configurations corresponding to a cluster of $n_{1}$ vortices
in the first band and $n_{2}$ vortices in the second band, out of which
$n_{c}$ vortices coincide.

In the present treatment, we do not consider superfluids corresponding to two
bands rotating relatively to each other, because the rotation frequency is
determined by a common stirring field. Therefore the vortex clusters in
different bands are in rest with respect to each other, forming a stable
configuration. The relative positions between vortices in two bands (both
radial and angular) are determined minimizing the free energy with respect to
positions of all vortices.

\begin{figure}
[th]
\begin{center}
\includegraphics[
height=6.6537in,
width=5.068in
]%
{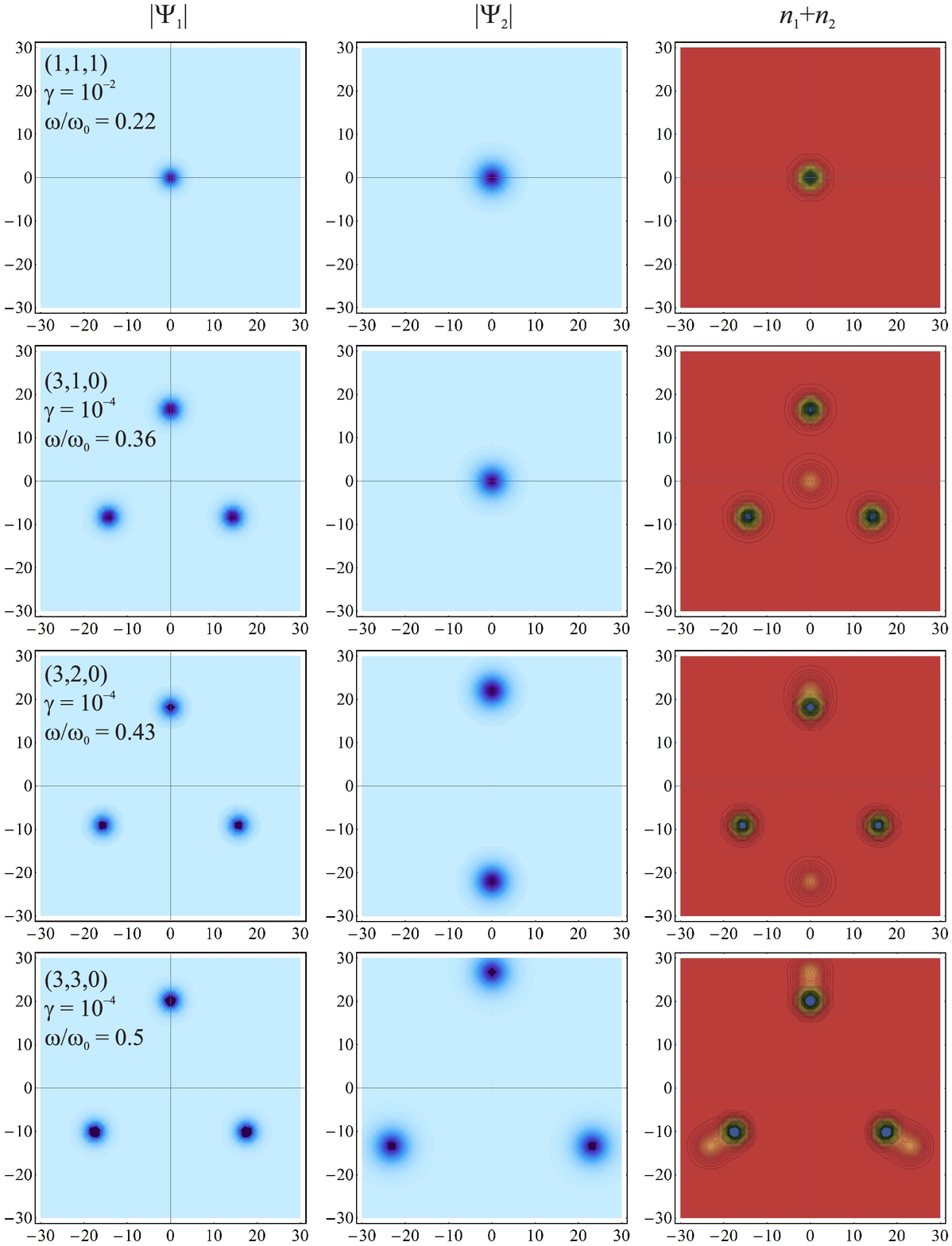}%
\caption{Absolute value of the pair field for several vortex states in the
first and second band-components of the condensate in a rotating Fermi gas
with the parameters $1/\left(  k_{F}a_{1,s}\right)  =0$, $1/\left(
k_{F}a_{2,s}\right)  =-0.5$, $T=0.01T_{F}$. The third column shows the total
particle density $n=n_{1}+n_{2}$. The values of the rotation frequency and the
interband coupling strength are explicitly given in the figure. The
coordinates are measured in units of the inverse to the Fermi wave number,
$1/k_{F}$.}%
\label{fig:Vortices}%
\end{center}
\end{figure}

Several resulting vortex clouds are exemplified in Fig. \ref{fig:Vortices}.
Analogous clouds for a two-band superconductor can be found, e. g., in Ref.
\cite{M2012}. The figure shows the spatial distribution of the amplitude of
the pair field $\left\vert \Psi_{j}\left(  \mathbf{r}\right)  \right\vert $ in
the \textquotedblleft strong\textquotedblright\ and \textquotedblleft
weak\textquotedblright\ band-components of the condensate for several stable
vortex states, both integer and fractional ones. The third column of Fig.
\ref{fig:Vortices} shows the spatial behavior of the total particle density
$n=n_{1}+n_{2}$. The particle density is necessarily modulated in vortices due
to the variation of the gap parameters. The modulation of the particle density
is experimentally observable. As can be seen from Fig. \ref{fig:Vortices},
fractional vortex configurations can be experimentally resolved from integer
vortex states due to a difference between vortex patterns in the two bands. In
the present work, we assume that the size of the trap is much larger than the
healing lengths of vortices in both band-components. This favors non-composite
vortices, except for the case when they are positioned in the center of the
trap. Therefore, for the vortex states shown in Fig. \ref{fig:Vortices}, only
the integer $\left(  1,1,1\right)  $ state is composite, and the other states
are non-composite. It is worth noting however that the distinction between
composite and non-composite vortices depends on convention. The interband
coupling results in an attraction between vortices in different
band-components of the condensate. This attraction tends to minimize a
non-compensated phase shift between the pair fields of the two
band-components. It leads to the binding of vortices in two bands. When the
distance between vortex cores in the two band-components of the condensate is
of the same order as the healing lengths (or less), it may become difficult to
resolve experimentally. In experiments, such bound vortices can be observed as
elongated. In this case, they are non-composite formally, but close to
composite vortices from the experimental point of view.

\subsection{Vortex phase diagrams}

\subsubsection{Phase diagrams in variables $\left(  \gamma,\omega\right)  $}

Analyzing the areas of stability for vortex configurations, we consider first
the evolution of the vortex states as a function of the interband coupling
strength $\gamma$. When the interband coupling is sufficiently strong, only
integer vortex states can survive, because the Josephson coupling contribution
to the free energy is positive and proportional to the relative phase
difference for the two band-components of the condensate. Therefore we are
particularly focused on the range of relatively small $\gamma$, where stable
fractional vortices can exist.

\begin{figure}
[ptbh]
\begin{center}
\includegraphics[
height=4.2575in,
width=6.4541in
]%
{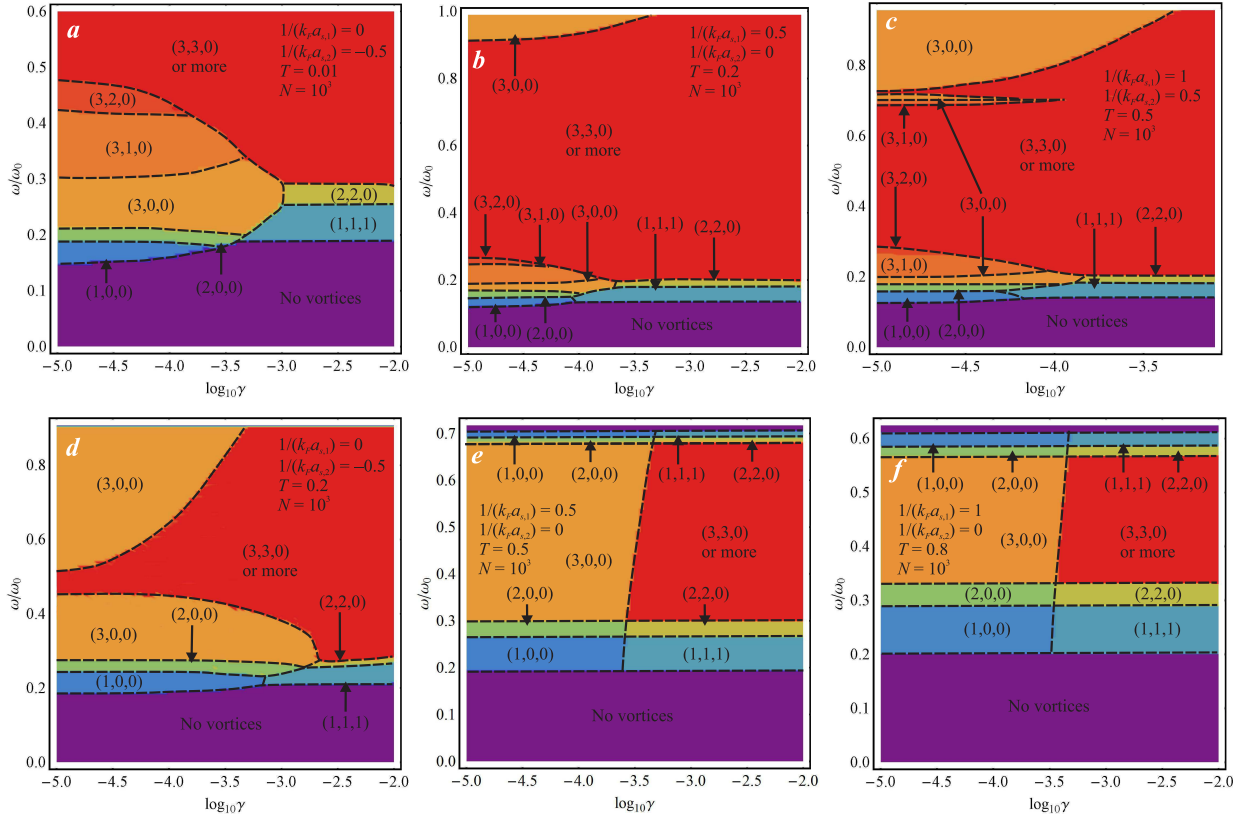}%
\caption{Vortex phase diagrams in the $\left(  \log_{10}\gamma,\omega\right)
$ variable space for a two-band rotating Fermi gas with the numbers of
fermions per unit length $N_{1}=N_{2}=500$.}%
\label{fig:Figure2}%
\end{center}
\end{figure}

\begin{figure}
[ptbh]
\begin{center}
\includegraphics[
height=4.3223in,
width=6.4515in
]%
{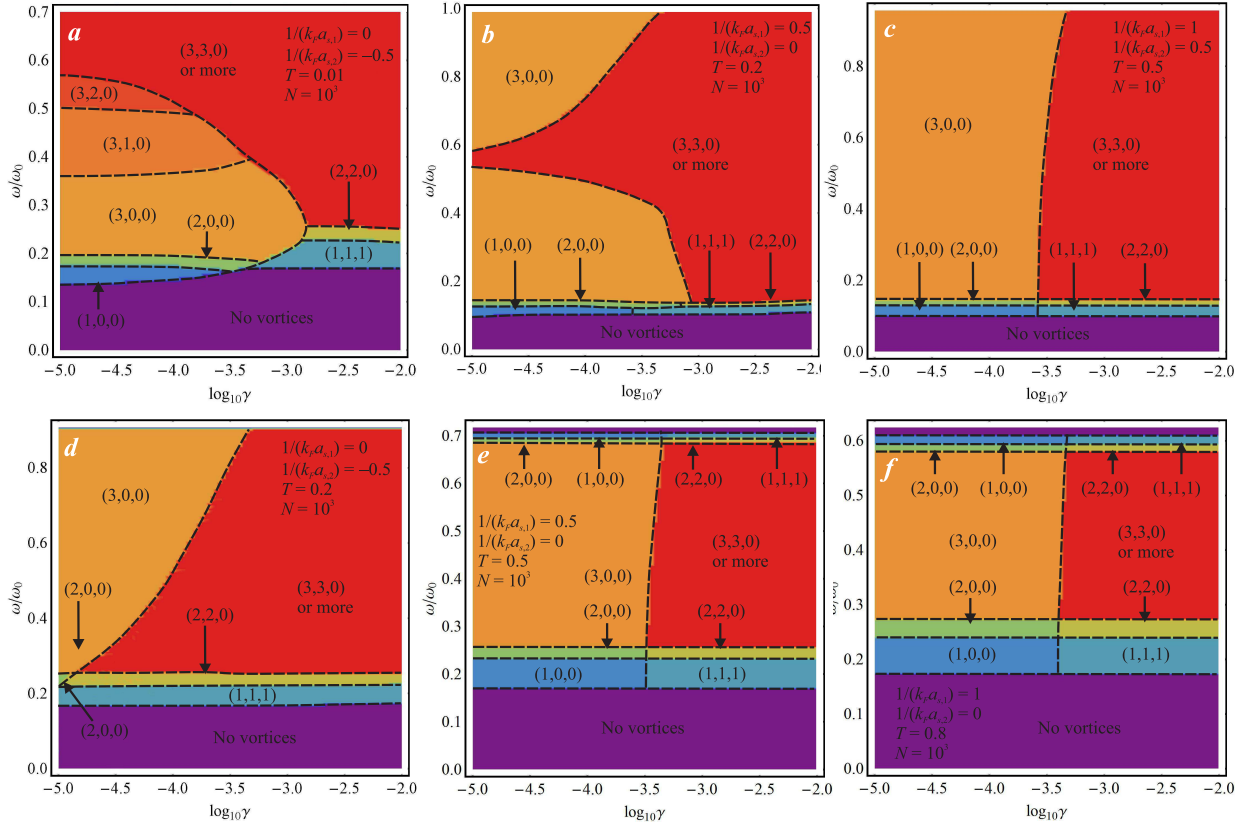}%
\caption{Vortex phase diagrams in the $\left(  \log_{10}\gamma,\omega\right)
$ variable space for a two-band rotating Fermi gas with the number of fermions
per unit length $N=10^{3}$, in an ensemble where $N_{1}$ and $N_{2}$ are fixed
by a common chemical potential.}%
\label{fig:Figure3}%
\end{center}
\end{figure}

In order to study this evolution of stable vortex configuration when varying
$\gamma$, we plot in Figs. \ref{fig:Figure2} and \ref{fig:Figure3} the vortex
phase diagrams in the variable space $\left(  \log_{10}\gamma,\omega\right)
$. The logarithmic scale for $\gamma$ is chosen because of our particular
attention to the range of weak interband couplings to track the transitions
between fractional and integer vortices. Fig. \ref{fig:Figure2} shows the
areas of stability for vortex states assuming numbers of particles per unit
length in each band fixed, $N_{1}=N_{2}=500$. Fig. \ref{fig:Figure3}
represents the analogous picture for the grand canonical ensemble of fermions
with the total number of particles per unit length $N=10^{3}$.

In Figs. \ref{fig:Figure2} and \ref{fig:Figure3}, when varying $\gamma$ at
fixed temperatures, a rich variety of stable vortex configurations, both
integer and fractional, can be seen in the vortex phase diagrams. The sequence
of stable vortex configurations with an increasing interband coupling is
physically quite transparent: fractional vortices turn to integer ones at a
critical coupling strength which only slightly depends on the rotation
frequency. On the contrary, when the \textquotedblleft weak\textquotedblright%
\ band is in the BCS regime, the dependence of the critical interband coupling
strength on the rotation frequency is rather complicated. As one can see from
Figs. \ref{fig:Figure2} and \ref{fig:Figure3} [especially clearly in panels
(\emph{e}, \emph{f}), but also in other phase diagrams], each vortex
configuration except $\left(  3,3,0\right)  $ may appear twice, at low and
high rotation frequencies. This trend in the sequences of the vortex stability
areas for $\left(  \gamma,\omega\right)  $ phase diagrams is related to the
fact that each \textquotedblleft phase\ boundary\textquotedblright\ in the
$\left(  T,\omega\right)  $ phase diagrams exhibits a bend-over dependence on
$\omega$. This bendover dependence is manifested in Figs. \ref{fig:Figure2}
and \ref{fig:Figure3} through an appearance of two distinct areas for given
vortex configurations (except when the upper-frequency areas lie in close
vicinity to the maximal rotation frequency $\omega_{\max}\left(  T\right)  $
and therefore are not seen). Note that the range of $\gamma$ for which
fractional vortices can exist is rather small, and that the vortex phase
diagrams in Figs. \ref{fig:Figure2} and \ref{fig:Figure3} are plotted in
logarithmic scale for $\gamma$. This explains the result that the phase
boundaries weakly depend on $\gamma$ in this range.

In the case of sufficiently weak interband coupling, the fractional vortices
are energetically more favorable with respect to integer vortices, except
maybe the low-temperature and high-rotation-frequency areas, where the
configuration $\left(  3,3,0\right)  $ and higher (in our notations, this
denotes three integer non-composite vortices) is energetically more favorable
with respect to fractional vortex states. This result does not ensure however
that the three-vortex integer configuration cannot be suppressed by a
fractional configuration of a higher order if we will account for more
vortices. When the interband coupling is a bit stronger, both fractional and
integer vortex configurations can survive. For a still stronger interband
coupling, we can observe only integer vortices and vortex systems. As a rule,
stable fractional vortex configurations appear in such a way that vortices
survive in the \textquotedblleft strong\textquotedblright\ band, when the
amplitude of the order parameter in the \textquotedblleft
weak\textquotedblright\ band becomes sufficiently small.

\subsubsection{Phase diagrams in variables $\left(  T,\omega\right)  $}

\begin{figure}
[ptbh]
\begin{center}
\includegraphics[
height=4.2531in,
width=6.3685in
]%
{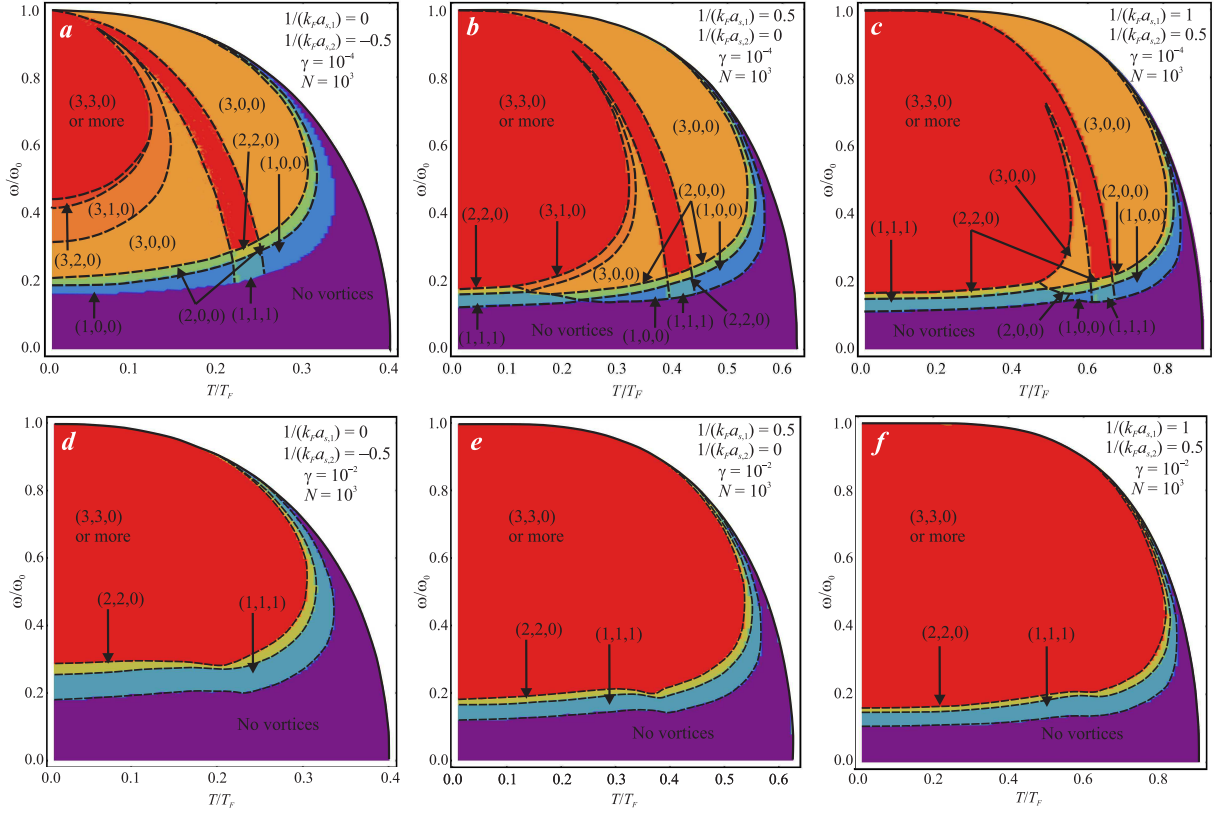}%
\caption{Vortex phase diagrams in the $\left(  T,\omega\right)  $ variable
space for a two-band rotating Fermi gas with different inverse scattering
lengths and interband coupling strengths indicated in the figure, for the
number of fermions per unit length $N=10^{3}$ ($N_{1}=N_{2}=500$).}%
\label{fig:Figure4}%
\end{center}
\end{figure}

\begin{figure}
[ptbh]
\begin{center}
\includegraphics[
height=4.2592in,
width=6.3045in
]%
{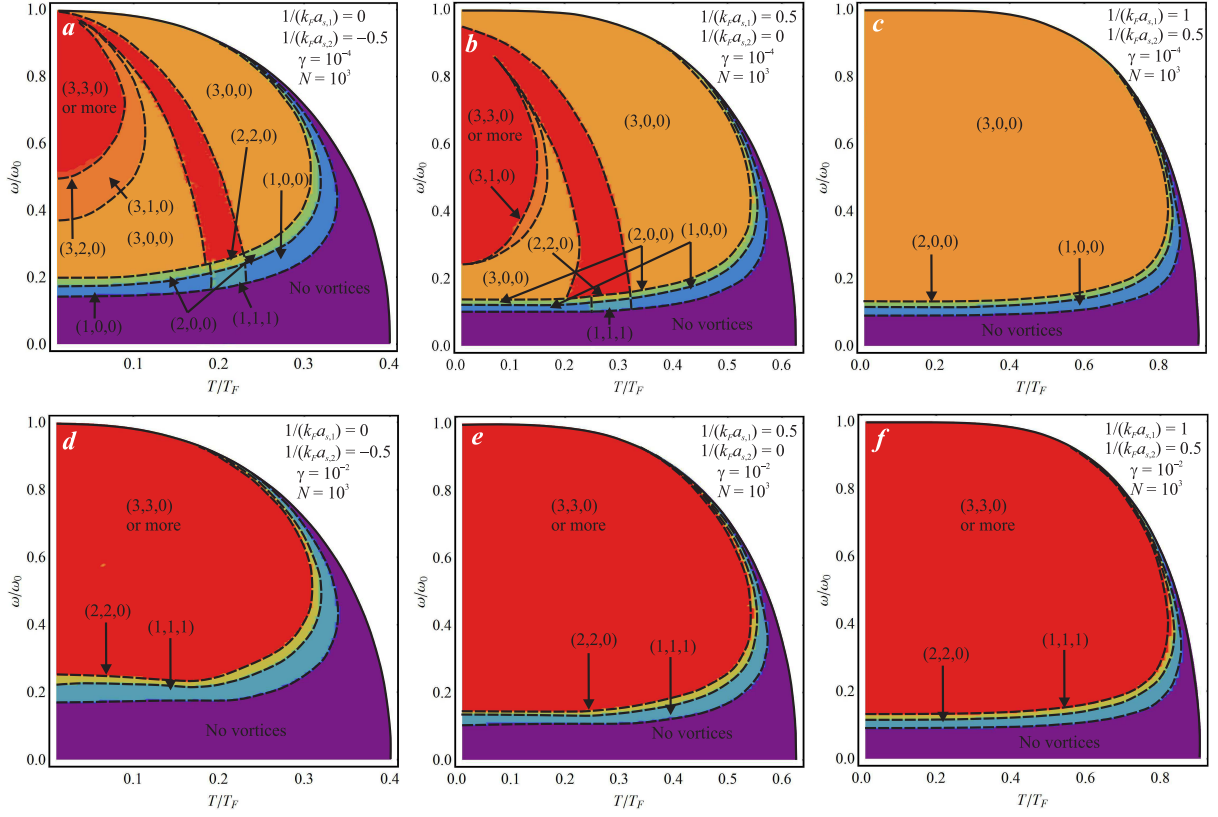}%
\caption{Vortex phase diagrams in the $\left(  T,\omega\right)  $ variable
space for a two-band rotating Fermi gas with different inverse scattering
lengths and interband coupling strengths indicated in the figure, for the
number of fermions per unit length $N=N_{1}+N_{2}=10^{3}$ in the
\textquotedblleft grand canonical\textquotedblright\ case when $N_{1}$ and
$N_{2}$ are determined by a common chemical potential.}%
\label{fig:Figure5}%
\end{center}
\end{figure}

Next, we analyze the vortex phase diagrams in the variables $\left(
T,\omega\right)  $ which show areas of stability for different vortex states
in a two-band rotating Fermi gas with several inverse scattering lengths and
different interband coupling strengths. Fig. \ref{fig:Figure4} shows the
vortex phase diagrams for the ensemble with the fixed numbers of particles per
unit length (along the $z$ direction) $N_{1}=N_{2}=500$. In Fig.
\ref{fig:Figure5}, the same fermionic system with $N\equiv N_{1}+N_{2}=1000$
is considered for the thermalized state with a common chemical potential. In
these figures, panels (\emph{a})-(\emph{c}) show vortex states at a rather
small interband coupling strength $\gamma=10^{-4}$. The panels (\emph{d}%
)-(\emph{f}) represent the vortex phase diagrams for a larger (but still
rather small) interband coupling $\gamma=10^{-2}$. The left-hand panels
(\emph{a,d}) of Figs. \ref{fig:Figure4} and \ref{fig:Figure5} correspond to
the case when the \textquotedblleft strong\textquotedblright\ band is at
unitarity, and \textquotedblleft weak\textquotedblright\ band is in the BCS
regime -- with, respectively, $1/a_{s,1}=0$ and $1/a_{s,2}=-0.5$. The central
panels (\emph{b,e}) show the case of stronger couplings, with, $1/a_{s,1}=0.5$
for the \textquotedblleft strong\textquotedblright\ band and $1/a_{s,2}=0$ for
the \textquotedblleft weak\textquotedblright\ band. Finally, the right-hand
panels (\emph{c,f}) show vortex phase diagrams for $1/a_{s,1}=1$ for the
\textquotedblleft strong\textquotedblright\ band and $1/a_{s,2}=0$ for the
\textquotedblleft weak\textquotedblright\ band, i. e., when the
\textquotedblleft strong\textquotedblright\ band is in the BEC regime.

As can be seen from Figs. \ref{fig:Figure2} and \ref{fig:Figure3}, the value
$\gamma=10^{-4}$ represents the specific interesting case, because it lies in
a rather narrow range of $\gamma$ where fractional vortices turn to integer
vortices. The other value, $\gamma=10^{-2}$, is chosen to show phase diagrams
where only integer vortex configurations survive. Higher interband coupling
strengths do not lead to substantial qualitative changes of phase diagrams,
because fractional vortices can be energetically favorable only at weak
interband coupling. This explains the choice of parameters made for Figs.
\ref{fig:Figure4} and \ref{fig:Figure5}.

In the same way as for a one-band Fermi gas, the vortex phase diagrams are
restricted to the area of the superfluid state, i.~e., to temperatures $T$
below the critical temperature $T_{c}\left(  \omega\right)  $. The critical
temperature is frequency-dependent and tends to zero when $\omega$ tends to
$\omega_{0}$. The value $\gamma=10^{-4}$ for a weak interband coupling is
chosen here because, on the one hand, it is sufficiently small so that the
vortex phase diagrams can contain fractional vortex states, and, on the other
hand, this value, although being small, is sufficient to yield rich stability
areas for both fractional and integer vortex states. The other set of phase
diagrams has been calculated for a relatively small but larger interband
coupling strength $\gamma=10^{-2}$, at which only integer vortex states
survive. Similarly to vortex phase diagrams for a one-band Fermi gas
\cite{Warringa2,Simonucci2015,RotVort1}, the boundaries of the areas of
stability for different vortex configurations exhibit a non-monotonic
dependence on the rotation frequency and a reentrant behavior as a function of
the temperature. The explanation of this bendover behavior of critical
rotation frequencies is the same for two-band and one-band Fermi gases. As
discussed in Refs. \cite{Warringa2,RotVort1}, it is related to a decrease of
the radius of the superfluid core for a trapped Fermi gas when the rotation
frequency becomes close to the confinement frequency. The lower critical
rotation frequency corresponds to the ordinary threshold for vortex formation,
similarly to that in a superconductor in an external magnetic field, when
increasing the field strength. The other (upper) critical rotation frequency
appears when the radius of the superfluid core is comparable with the healing length.

When comparing the vortex phase diagrams for ensembles with fixed numbers of
particles and with the common chemical potential to each other, one can see
that the \textquotedblleft grand canonical\textquotedblright\ ensemble is more
favorable for fractional vortices than that with fixed numbers of particles.
This difference is explained by the effect of a partial depletion of a
\textquotedblleft weak\textquotedblright\ band. As found in Ref.
\cite{KTLD2015}, this depletion is a feedback of the gap parameter to the
relative band populations. As a result, the areas of stability for fractional
vortices in the case of a common chemical potential (Fig. \ref{fig:Figure5})
are wider than in the case of fixed equal numbers of particles in each band.
Moreover, as can be seen comparing Fig. \ref{fig:Figure4} (\emph{c}) and
\ref{fig:Figure5} (\emph{c}), in the BEC regime this depletion can lead to a
complete vanishing of integer vortex phases.

\begin{figure}
[ptbh]
\begin{center}
\includegraphics[
height=3.4368in,
width=4.5394in
]%
{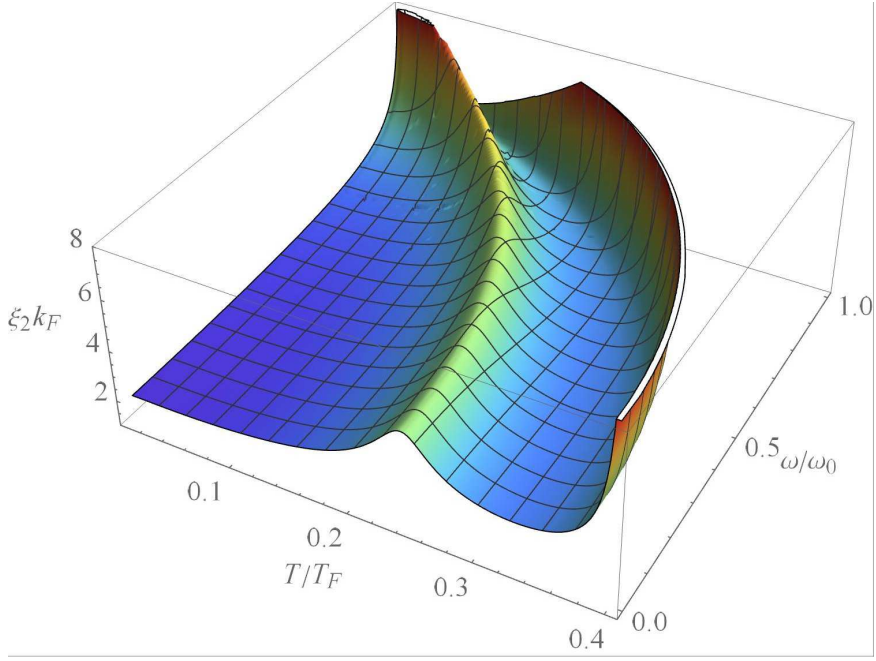}%
\caption{Healing length for a \textquotedblleft weak\textquotedblright\ band
$\xi_{2}$ in a two-band rotating Fermi gas with the inverse scattering lengths
$1/\left(  k_{F}a_{s,1}\right)  =0,\;1/\left(  k_{F}a_{s,2}\right)  =-0.5$,
the numbers of fermions per unit length $N_{1}=N_{2}=500$, and the interband
coupling strength $\gamma=10^{-2}$ as a function of the temperature and the
rotation frequency.}%
\label{fig:Figure6}%
\end{center}
\end{figure}

Distinct manifestations of the interband coupling in a two-band Fermi gas are
related to the temperature $T_{c,2}$ equal to the critical temperature of the
superfluid phase transition in the \textquotedblleft weak\textquotedblright%
\ band-condensate component in the absence of the interband coupling. First,
we find a kink of the \textquotedblleft phase boundaries\textquotedblright%
\ between different vortex configurations in Figs. \ref{fig:Figure4} and
\ref{fig:Figure5}, positioned at $T_{c,2}$. This feature is absent in the
one-band system considered in Refs. \cite{Warringa2,Simonucci2015,LAKT2015}.
It is directly related to a non-monotonic, resonant peak-shape temperature
dependence of the vortex healing length for the \textquotedblleft
weak\textquotedblright\ band $\xi_{2}$, analyzed in Refs.
\cite{KTLD2015,Babaev2012-2,KomendovaPRL108} and termed \textquotedblleft
hidden criticality\textquotedblright\ in Ref. \cite{KomendovaPRL108}. Although
there is no true criticality at $T\approx T_{c,2}$, many parameters of state
demonstrate non-monotonic behavior at this temperature. The peak position for
$\xi_{2}$ lies at the critical temperature for the \textquotedblleft
weak\textquotedblright\ band in the absence of the interband coupling, so that
it is a fingerprint of the weak-band criticality, which is lost in a two-band
system. In Fig. \ref{fig:Figure6}, the evolution of this \textquotedblleft
hidden criticality\textquotedblright\ point is shown as a function of $\left(
T,\omega\right)  $ for the inverse scattering lengths $1/\left(  k_{F}%
a_{s,1}\right)  =0,\;1/\left(  k_{F}a_{s,2}\right)  =-0.5$, the numbers of
fermions per unit length $N_{1}=N_{2}=500$, and the interband coupling
strength $\gamma=10^{-2}$. The peak temperature of the healing length $\xi
_{2}$ diminishes when the rotation frequency rises, with an increasing peak
magnitude and a decreasing width.

The seemingly unusual sequence of phases in the phase diagrams for weak
interband coupling (upper panels of Figs. \ref{fig:Figure4} and
\ref{fig:Figure5}) can be transparently understood using simple physical
reasoning. First, let us temporarily \textquotedblleft turn
off\textquotedblright\ the interband interaction and consider each
band-component independently as a one-band system. For a one-band trapped
Fermi gas, this behavior was independently analyzed in Refs.
\cite{Warringa2,Simonucci2015,RotVort1}, using different methods: Bogoliubov
-- de Gennes theory and EFT. These two theories completely agree in
predictions on the behavior of vortex states.

The evolution of the number of vortices in a given band-condensate, as a
function of the rotation rate at a fixed temperature is non-monotonic. The
healing length, which is a parameter characterizing the vortex size, is an
increasing function of the rotation frequency. As established in Refs.
\cite{Warringa2,Simonucci2015,RotVort1}, the number of vortices increases when
rotation frequency increases as long as the vortex size remains substantially
smaller than the size of the superfluid core. When the rotation frequency is
further increased, an upper critical rotation frequency appears such that the
vortex state turns back to the superfluid state for higher rotation
frequencies, because the radius of the superfluid state reduces when the
rotation frequency becomes close to the maximal rotation frequency at which
superfluidity vanishes. It was also found in Ref. \cite{RotVort1} that the
boundary between the states with different numbers of vortices behaves
similarly to the critical rotation frequency for a single vortex. For example,
the lower critical rotation frequency for a vortex pair is higher than the
lower critical rotation frequency for a single vortex. On the contrary, the
upper critical rotation frequency for a vortex pair is lower than the upper
critical rotation frequency for a single vortex.

Regarding the variation of the number of vortices as a function of temperature
at a finite rotation rate (sufficient for stable vortices at $T=0$), the
number of vortices monotonically decreases with increasing temperature,
because the free energy of the system with vortices becomes higher when the
vortex size increases (as an increasing function of temperature). For a
two-band system without interband coupling, similarly to the one-band system,
the number of vortices in each band-component of the condensate must be a
monotonically decreasing function of temperature. Therefore vortices in the
\textquotedblleft weak\textquotedblright\ band in the absence of the interband
coupling cannot exist above the critical temperature $T_{c,2}<T_{c,1}$, while
vortices in the \textquotedblleft strong\textquotedblright\ band can survive
above $T_{c,2}$ and vanish at higher temperatures close to $T_{c,1}$, as can
be seen from Figs. \ref{fig:Figure4} and \ref{fig:Figure5}).

The monotonic dependence of vortex numbers as a function of temperature can be
broken by interband coupling, because in the weakly-coupled two-band system
the healing length for the \textquotedblleft weak\textquotedblright%
\ band-component exhibits a pronounced peak near $T_{c,2}$, as shown in
\cite{KTLD2015}. In particular, the healing length for the \textquotedblleft
weak\textquotedblright\ band reaches a local maximum at $T\approx T_{c,2}$ and
again reduces when temperature slightly exceeds $T_{c,2}$. This favors an
increasing number of vortices in the \textquotedblleft weak\textquotedblright%
\ band-component of the condensate in a relatively narrow temperature area
above $T_{c,2}$. Since the effects related to \textquotedblleft hidden
criticality\textquotedblright\ are very transparent and necessarily follow
from the interband coupling, which is a highly controllable parameter (see,
e.~g., \cite{Zhang2015,Xu2016}), the sequence of phases obtained in our work
can be experimentally observable and serve as a clear \textquotedblleft
smoking gun\textquotedblright\ evidence for multiband physics in superfluids.

\begin{figure}
[ptbh]
\begin{center}
\includegraphics[
height=4.0179in,
width=4.9554in
]%
{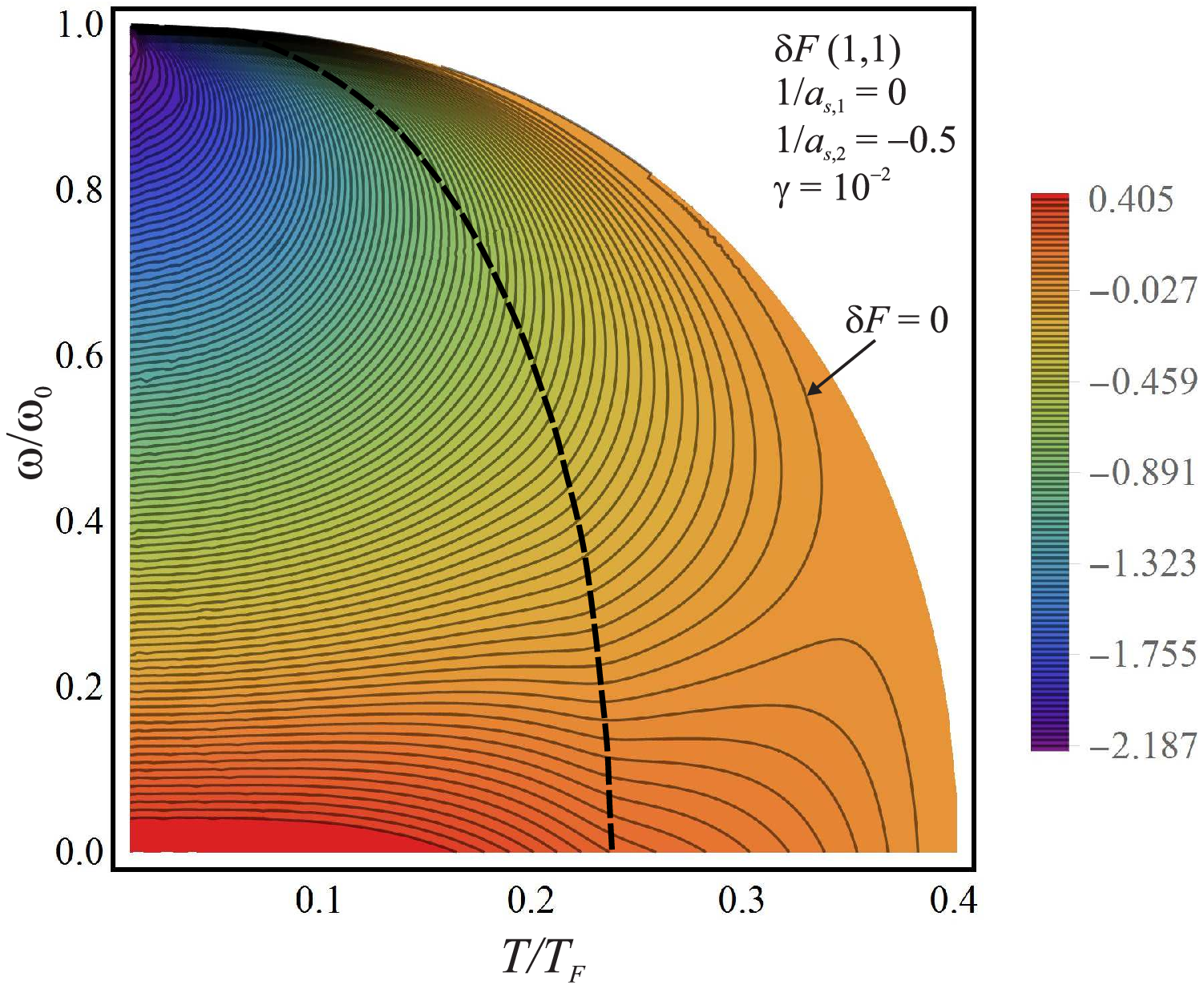}%
\caption{Free energy for an integer vortex state $\left(  1,1,1\right)  $ in a
two-band rotating Fermi gas with the inverse scattering lengths $1/\left(
k_{F}a_{s,1}\right)  =0,\;1/\left(  k_{F}a_{s,2}\right)  =-0.5$, the numbers
of fermions per unit length $N_{1}=N_{2}=500$, and the interband coupling
strength $\gamma=10^{-2}$ as a function of the temperature and the rotation
frequency. The path of the peak for the healing length $\xi_{2}$ is shown by
the dashed curve.}%
\label{fig:Figure7}%
\end{center}
\end{figure}

The resonant dependence of the healing length for the \textquotedblleft
weak\textquotedblright\ band influences the free energy of the two-band Fermi
gas, as can be seen from Fig. \ref{fig:Figure7}. This figure shows the free
energy difference $\delta F$ for an integer vortex, denoted as $\delta
F\left(  1,1\right)  $ for $1/\left(  k_{F}a_{s,1}\right)  =0,\;1/\left(
k_{F}a_{s,2}\right)  =-0.5,$ and $\gamma=10^{-2}$. Thus both the plot for the
healing length in Fig. \ref{fig:Figure6} and the free energy shown in Fig.
\ref{fig:Figure7} correspond to the same values of parameters as the vortex
phase diagram in Fig. \ref{fig:Figure4} (\emph{d}). The area $\left(
T,\omega\right)  $ for $T<T_{c}\left(  \omega\right)  $ can be subdivided to
two areas by the contour indicating $\delta F=0$ shown in the figure
explicitly. These areas correspond to $\delta F<0$ and $\delta F>0$
(respectively, inside and outside the contour for $\delta F=0$). As can be
seen from Fig. \ref{fig:Figure7}, the behavior of contour lines for the free
energy in these two areas is quite different. Also the free energy exhibits a
non-monotonic behavior of isoenergetic contours which contain kinks along the
path of the peak for the healing length $\xi_{2}$. The same kinks are
manifested in the contour lines of the vortex phase diagrams.

Second, the peak-shape dependence of the healing length in the
\textquotedblleft weak\textquotedblright\ band-component of the condensate
$\xi_{2}$ on the temperature leads to the fact that $\xi_{2}$ decreases in a
certain temperature range above $T_{c,2}$ to values of the same order as
$\xi_{1}$. The small healing length $\xi_{2}$ facilitates vortex stabilization
in the \textquotedblleft weak\textquotedblright\ band-component of the
condensate. As a result, as can be seen from Figs. \ref{fig:Figure4}
(\emph{a}, \emph{b}, \emph{c}) and \ref{fig:Figure5} (\emph{a}, \emph{b}), an
area of integer vortex states appears above $T_{c,2}$ [see, e. g., Fig.
\ref{fig:Figure6} which corresponds to the same set of parameters as Fig.
\ref{fig:Figure4} (\emph{a})]. Note that this area of integer vortex states
above $T_{c,2}$ is a consequence of a non-zero interband coupling, because at
$\gamma=0$ there is no condensate in the \textquotedblleft
weak\textquotedblright\ band. The appearance of this resonant regime is not
present in vortex phase diagrams for a one-band Fermi condensate. For
comparison, in a one-band system, the sequence of different vortex
configurations is such that the areas of stability for higher winding numbers
lie completely inside the areas of stability for lower winding numbers
\cite{Warringa2,RotVort1} in phase diagrams. For a two-band system, this
ordering of vortex states can be violated. Namely, at $T<T_{c,2}$ we observe
the usual\ sequence of stability areas, where integer vortex configurations
change to fractional states with an increasing temperature. For $T>T_{c,2}$,
the \textquotedblleft anomalous\textquotedblright\ sequence becomes possible,
where fractional states change to integer states when temperature rises. When
temperature further increases towards $T_{c}$, the sequence of stability areas
becomes usual\ again. The appearance of the \textquotedblleft
anomalous\textquotedblright\ sequence of stability areas for fractional and
integer vortex states is arguably the most striking result of the interband
coupling for two-band rotating Fermi gases.

\subsection{Relevance to experiments}

Recently, a new kind of orbital Feshbach resonance has been experimentally
achieved in ultracold $^{173}$Yb fermionic atoms \cite{Pagano2015,Hofer2015}.
As shown in Refs. \cite{Iskin2016,Iskin2017,He}, the many-body Hamiltonian
describing this system with an orbital Feshbach resonance is exactly analogous
to that of a two-band $s$-wave fermionic superfluid with contact interaction.
We can map the parameters corresponding to these experiments on the present
theory. According to Refs. \cite{Pagano2015,Hofer2015,Iskin2016}, the
orbital-singlet and orbital-triplet scattering lengths are, respectively,
$a_{s+}\approx1900a_{0}$ and $a_{s-}\approx200a_{0}$, where $a_{0}$ is the
Bohr radius. In the experimental conditions of Ref. \cite{Pagano2015}, an
ultracold gas of $N=6\times10^{4}$ fermionic $^{173}$Yb atoms is confined to a
cigar-like trap with $\omega_{x}=2\pi\times13%
\operatorname{Hz}%
$, $\omega_{y}=2\pi\times188%
\operatorname{Hz}%
$, and $\omega_{z}=2\pi\times138%
\operatorname{Hz}%
$. This gives us the Fermi energy $E_{F}=\hbar\left(  3\omega_{x}\omega
_{y}\omega_{z}N\right)  ^{1/3}\approx2.604\times10^{-30}%
\operatorname{J}%
$, and the parameter $k_{F}\approx1.1597\times10^{7}%
\operatorname{m}%
^{-1}$. Using the set of units described in Sec. II and following to Ref.
\cite{Iskin2016}, we find the dimensionless parameters corresponding to the
experiment \cite{Pagano2015} to be $1/a_{s,1}=1/a_{s,2}\approx4.50$, and
$\gamma\approx3.64$. This strong interband coupling only allows for integer vortices.

As can be seen from Figs. 2 and 3, in the range $\gamma\sim10^{-3}$ to
$\gamma\sim10^{-2}$, where only integer vortex states survive, the boundaries
between areas of stability for different vortex states depend on $\gamma$
extremely weakly. Therefore the same sequence of vortex states when varying
temperature as in Figs. 4 (d, e, f) and 5 (d, e, f) must be observed also at
stronger interband couplings, which are relevant for the experimentally
realized case of $^{173}$Yb atoms \cite{Pagano2015,Hofer2015}.

The scattering lengths are highly controllable for ultracold Fermi gases. One
can therefore expect that low values of the interband coupling strength, which
are needed for fractional vortices in multiband superfluids, are realizable in
future experiments.

\section{Conclusions \label{sec:Conclusions}}

In this work, we have applied the finite temperature effective field theory to
investigate integer and fractional vortices and multivortex states in rotating
two-band Fermi gases throughout the BCS-BEC crossover. As distinct from the
one-band system, a rich spectrum of vortex states is realizable in a two-band
Fermi gas. Fractional vortices (the states with different winding numbers in
the two band-components of the condensate) are stable in this system for
sufficiently weak interband couplings. When the interband coupling strength
$\gamma$ exceeds a critical value, which is dependent on the frequency of
rotation $\omega$, only integer vortices and vortex clusters can be found.
Note that integer vortices in many-vortex clusters are not necessarily
composite: the vortex centers in \textquotedblleft strong\textquotedblright%
\ and \textquotedblleft weak\textquotedblright\ band may reside at different
distances from the trap center.

The phase boundaries\ between different stable vortex configurations in the
$\left(  T,\omega\right)  $ vortex phase diagrams depend non-monotonically on
the rotation frequency, turning to zero both at lower and upper critical
rotation frequencies at $T=0$. Correspondingly, they exhibit a bend-over
temperature dependence, quite similar to those obtained in preceding works for
a one-band Fermi gas.

We have characterized the difference between vortex phase diagrams obtained
for two-band Fermi condensates in two different regimes: the regime of fixed
numbers of particles for each band-components of the condensate, and the grand
canonical ensemble when these numbers of particles are determined through the
common chemical potential. The depletion of the \textquotedblleft
weak\textquotedblright\ band-component of the condensate in the grand
canonical ensemble can lead to a substantial expansion of stability areas with
respect to those at fixed numbers of particles. This difference can be
experimentally verified.

A striking difference appears between the evolution of vortex configurations
as a function of $\gamma$ in the BCS and BEC regimes. A rich variety of
fractional vortex states exist at weak interband couplings in a wide range of
rotational frequencies, which turn to integer vortex states at certain
$\gamma$. The boundaries between different vortex states strongly depend on
$\gamma$ in the BCS regime. On the contrary, in the BEC regime, these
boundaries rather weakly depend on the interband coupling strength.

The obtained manifestations of the non-monotonic behavior of the healing
length in the \textquotedblleft weak\textquotedblright\ band-component of the
condensate through kinks of phase boundaries and the \textquotedblleft
anomalous\textquotedblright\ resonant sequence of stability areas in vortex
phase diagrams can be experimentally accessible. Usually, the
\textquotedblleft hidden criticality\textquotedblright\ phenomena and other
effects of the interband coupling cannot be easily resolved experimentally.
Therefore, the aforesaid manifestations can represent an effective and
transparent \textquotedblleft smoking gun\textquotedblright\ for the interband
coupling in trapped atomic Fermi superfluids.

\begin{acknowledgments}
We thank C. A. R. S\'{a} de Melo and N. Verhelst for valuable discussions.
This work has been supported by the Research Foundation-Flanders (FWO-Vl),
project nrs. G.0115.12N, G.0119.12N, G.0122.12N, G.0429.15N, G.0666.16N, by
the Scientific Research Network of the Flemish Research Foundation,
WO.033.09N, and by the Research Fund of the University of Antwerp.
\end{acknowledgments}

\appendix

\section{Gradient expansion for the effective bosonic action}

The exact effective bosonic action for interacting fermions is obtained
introducing auxiliary Hubbard-Stratonovich fields and integrating the
partition function out fermion fields as described in Ref. \cite{KTLD2015}.
The resulting effective bosonic action for a two-band system reads,%
\begin{equation}
S_{eff}^{\left(  2b\right)  }=\sum_{j=1,2}S_{eff}^{\left(  j\right)  }%
-\int_{0}^{\beta}d\tau\int d\mathbf{r~}\frac{\sqrt{m_{1}m_{2}}\gamma}{4\pi
}\left(  \bar{\Psi}_{1}\Psi_{2}+\bar{\Psi}_{2}\Psi_{1}\right)  , \label{SEFT}%
\end{equation}
where $S_{eff}^{\left(  j\right)  }$ is the effective field action for a
single band-component of the condensate determined by (dropping the band index
$j$):%
\begin{equation}
S_{eff}=S_{B}-\operatorname{Tr}\ln\left[  -\mathbb{G}^{-1}\right]  .
\label{Seff1a}%
\end{equation}
We follow here the notations of Ref. \cite{deMelo1993}, where $\mathbb{G}%
^{-1}\left(  \mathbf{r},\tau\right)  =\mathbb{G}_{0}^{-1}\left(
\mathbf{r},\tau\right)  -\mathbb{F}\left(  \mathbf{r},\tau\right)  $ is the
inverse Nambu tensor. It is subdivided to the free-fermion inverse Nambu
tensor $\mathbb{G}_{0}^{-1}$ and the matrix $\mathbb{F}$ proportional to the
pair field $\Psi$:%
\begin{align}
\mathbb{G}_{0}^{-1}\left(  \mathbf{r},\tau\right)   &  =\left(
\begin{array}
[c]{cc}%
-\frac{\partial}{\partial\tau}-\hat{H}_{\uparrow} & 0\\
0 & -\frac{\partial}{\partial\tau}+\hat{H}_{\downarrow}%
\end{array}
\right)  ,\label{G0a}\\
\mathbb{F}\left(  \mathbf{r},\tau\right)   &  =\left(
\begin{array}
[c]{cc}%
0 & -\Psi\left(  \mathbf{r},\tau\right) \\
-\bar{\Psi}\left(  \mathbf{r},\tau\right)  & 0
\end{array}
\right)  . \label{Fa}%
\end{align}
Next, the effective action (\ref{Seff1a}) is expanded as a Taylor series in
powers of the pair field:%
\begin{equation}
S_{eff}=S_{B}-\operatorname{Tr}\ln\left[  -\mathbb{G}_{0}^{-1}\right]
+\sum_{p=1}^{\infty}\frac{1}{p}\operatorname{Tr}\left[  \left(  \mathbb{G}%
_{0}\mathbb{F}\right)  ^{p}\right]  . \label{Sp}%
\end{equation}
In more detail, the trace $\operatorname{Tr}\left[  \left(  \mathbb{G}%
_{0}\mathbb{F}\right)  ^{p}\right]  $ is written as:%
\begin{align}
\operatorname{Tr}\left[  \left(  \mathbb{G}_{0}\mathbb{F}\right)  ^{p}\right]
&  =\int d\tau_{1}\ldots d\tau_{p}\int d\mathbf{r}_{1}\ldots d\mathbf{r}%
_{p}\operatorname{Tr}\left[  \mathbb{F}\left(  \mathbf{r}_{1},\tau_{1}\right)
\mathbb{G}_{0}\left(  \mathbf{r}_{1}-\mathbf{r}_{2},\tau_{1}-\tau_{2}\right)
\right. \nonumber\\
&  \left.  \times\mathbb{F}\left(  \mathbf{r}_{2},\tau_{2}\right)
\mathbb{G}_{0}\left(  \mathbf{r}_{2}-\mathbf{r}_{3},\tau_{2}-\tau_{3}\right)
\ldots\mathbb{F}\left(  \mathbf{r}_{p},\tau_{p}\right)  \mathbb{G}_{0}\left(
\mathbf{r}_{p}-\mathbf{r}_{1},\tau_{p}-\tau_{1}\right)  \right]  .
\label{Trace}%
\end{align}
The free-fermion Green's function matrix $\mathbb{G}_{0}\left(  \mathbf{r}%
,\tau\right)  $ is expressed through the Fourier representation,%
\begin{subequations}
\begin{align}
\mathbb{G}_{0}\left(  \mathbf{r},\tau\right)   &  =\frac{1}{\beta V}%
\sum_{\mathbf{k},n}\mathbb{G}_{0}\left(  \mathbf{k},n\right)  e^{i\mathbf{k}%
\cdot\mathbf{r}-i\omega_{n}\tau},\label{G0}\\
\mathbb{G}_{0}\left(  \mathbf{k},n\right)   &  =\left(
\begin{array}
[c]{cc}%
\frac{1}{i\left(  \omega_{n}-i\zeta\right)  -\xi_{\mathbf{k}}} & 0\\
0 & \frac{1}{i\left(  \omega_{n}-i\zeta\right)  +\xi_{\mathbf{k}}}%
\end{array}
\right)  , \label{Fourier}%
\end{align}
with the free-fermion energies%
\end{subequations}
\begin{equation}
\xi_{\mathbf{k}}=\frac{k^{2}}{2m}-\mu, \label{en}%
\end{equation}
and the fermion Matsubara frequencies%
\begin{equation}
\omega_{n}=\frac{\left(  2n+1\right)  \pi}{\beta}. \label{wn}%
\end{equation}
For spin-imbalanced fermions, $\mu$ in (\ref{en}) is the averaged chemical
potential $\mu=\left(  \mu_{\uparrow}+\mu_{\downarrow}\right)  /2$. The
chemical potential imbalance parameter is $\zeta=\left(  \mu_{\uparrow}%
-\mu_{\downarrow}\right)  /2$.

The action (\ref{Sp}) is still exact. Further, various approximations are
possible. The crudest one is the saddle-point (mean-field) approximation,
where $\Psi\left(  \mathbf{r}_{n},\tau_{n}\right)  $ is replaced to a constant
value $\Delta$ which realizes the least action principle for $S_{eff}$. The
frequently used approach beyond the saddle-point approximation accounts for
Gaussian pair fluctuations about the saddle-point value \cite{deMelo1993}. It
assumes that fluctuations are small with respect to $\Delta$. We apply the
alternative method which does not use an assumption of small fluctuations but
is related to conditions when the pair fields slowly vary in time and space.
Many approximations (in particular, the Ginzburg-Landau and Gross-Pitaevskii
theories) use the same assumption. Within this scheme, the first-step
extension of the saddle-point approximation for fermions non-uniformly
distributed in space (e.~g., for fermions trapped to a confinement potential)
is the local field approximation (LFA). Within this approximation, we set the
space and time variables for each $\mathbb{F}\left(  \mathbf{r}_{n},\tau
_{n}\right)  $ in (\ref{Trace}) to be the same, e.~g., $\mathbb{F}\left(
\mathbf{r}_{n},\tau_{n}\right)  =\mathbb{F}\left(  \mathbf{r}_{1},\tau
_{1}\right)  $. This leads to the exact summation over $p$, resulting in the
LFA effective action,%
\[
S_{LFA}=\int_{0}^{\beta}d\tau\int d\mathbf{r}~\Omega_{s}\left(  \left\vert
\Psi\left(  \mathbf{r},\tau\right)  \right\vert ^{2}\right)
\]
where $\Omega_{s}\left(  \left\vert \Psi\left(  \mathbf{r},\tau\right)
\right\vert ^{2}\right)  $ has the same form as the saddle-point thermodynamic
potential but with the coordinate-dependent squared modulus of the order
parameter (and also with a coordinate-dependent chemical potential):%
\begin{align}
\Omega_{s}\left(  \left\vert \Psi\right\vert ^{2}\right)   &  =-\int
\frac{d\mathbf{k}}{\left(  2\pi\right)  ^{3}}\left[  \frac{1}{\beta}\ln\left(
2\cosh\beta\zeta+2\cosh\beta E_{\mathbf{k}}\right)  -\xi_{\mathbf{k}}%
-\frac{m\left\vert \Psi\right\vert ^{2}}{k^{2}}\right] \nonumber\\
&  -\frac{m\left\vert \Psi\right\vert ^{2}}{4\pi a_{s}}. \label{Ws}%
\end{align}
Here $E_{\mathbf{k}}=\sqrt{\xi_{\mathbf{k}}^{2}+\left\vert \Psi\right\vert
^{2}}$ is the Bogoliubov excitation energy.

The next step within this scheme beyond the saddle-point approximation is the
gradient expansion of the pair fields $\Psi\left(  \mathbf{r}_{n},\tau
_{n}\right)  ,\bar{\Psi}\left(  \mathbf{r}_{n},\tau_{n}\right)  $ about one
common point, e.~g., about $\left(  \mathbf{r}_{1},\tau_{1}\right)  $. There
is no difference which of the fields is chosen as the background, because the
trace (\ref{Trace}) is invariant with respect to cyclic permutations of pair
fields. The gradient expansion is then:
\begin{align}
\Psi\left(  \mathbf{r}_{n},\tau_{n}\right)   &  =\Psi\left(  \mathbf{r}%
_{1},\tau_{1}\right)  +(\tau_{n}-\tau_{1})\frac{\partial\Psi\left(
\mathbf{r}_{1},\tau_{1}\right)  }{\partial\tau_{1}}+\frac{1}{2}(\tau_{n}%
-\tau_{1})^{2}\frac{\partial^{2}\Psi\left(  \mathbf{r}_{1},\tau_{1}\right)
}{\partial\tau_{1}^{2}}\nonumber\\
&  +\left(  \mathbf{r}_{n}-\mathbf{r}_{1}\right)  \cdot\nabla\Psi\left(
\mathbf{r}_{1},\tau_{1}\right)  +\frac{1}{2}\sum_{i,j=1}^{3}(x_{n,i}%
-x_{1,i})(x_{n,j}-x_{1,j})\frac{\partial^{2}\Psi\left(  \mathbf{r}_{1}%
,\tau_{1}\right)  }{\partial x_{1,i}\partial x_{1,j}}+\mathbf{...}
\label{Fexp}%
\end{align}
and the same for the conjugated pair field. Here, we restrict this series up
to the second order derivatives in time and space. Obviously, the zeroth-order
term of the gradient expansion corresponds to the aforesaid LF approximation.
Also for other terms of (\ref{Fexp}) substituted in (\ref{Sp}), the whole sum
over $p$ is collected analytically (for each term of the gradient expansion
separately). As a result, we arrive at the EFT action functional derived in
Ref. \cite{KTLD2015} (denoting $w\equiv\left\vert \Psi\right\vert ^{2}$):%
\begin{align}
&  S_{EFT}=\int_{0}^{\beta}d\tau\int d\mathbf{r}\left[  \Omega_{s}\left(
w\right)  +\frac{D\left(  w\right)  }{2}\left(  \bar{\Psi}\frac{\partial\Psi
}{\partial\tau}-\frac{\partial\bar{\Psi}}{\partial\tau}\Psi\right)  \right.
\nonumber\\
&  +Q\left(  w\right)  \frac{\partial\bar{\Psi}}{\partial\tau}\frac
{\partial\Psi}{\partial\tau}-\frac{R\left(  w\right)  }{2w}\left(
\frac{\partial w}{\partial\tau}\right)  ^{2}\nonumber\\
&  \left.  +\frac{C\left(  w\right)  }{2m}\left(  \nabla_{\mathbf{r}}\bar
{\Psi}\cdot\nabla_{\mathbf{r}}\Psi\right)  -\frac{E\left(  w\right)  }%
{2m}\left(  \nabla_{\mathbf{r}}w\right)  ^{2}\right]  , \label{FGL2}%
\end{align}
with the coefficients:%
\begin{align}
C  &  =\int\frac{d\mathbf{k}}{\left(  2\pi\right)  ^{3}}\frac{k^{2}}{3m}%
f_{2}\left(  \beta,E_{\mathbf{k}},\zeta\right)  ,\label{c}\\
D  &  =\int\frac{d\mathbf{k}}{\left(  2\pi\right)  ^{3}}\frac{\xi_{\mathbf{k}%
}}{w}\left[  f_{1}\left(  \beta,\xi_{\mathbf{k}},\zeta\right)  -f_{1}\left(
\beta,E_{\mathbf{k}},\zeta\right)  \right]  ,\label{d}\\
E  &  =2\int\frac{d\mathbf{k}}{\left(  2\pi\right)  ^{3}}\frac{k^{2}}{3m}%
\xi_{\mathbf{k}}^{2}~f_{4}\left(  \beta,E_{\mathbf{k}},\zeta\right)
,\label{ee}\\
Q  &  =\frac{1}{2w}\int\frac{d\mathbf{k}}{\left(  2\pi\right)  ^{3}}\left[
f_{1}\left(  \beta,\xi_{\mathbf{k}},\zeta\right)  -\left(  E_{\mathbf{k}}%
^{2}+\xi_{\mathbf{k}}^{2}\right)  f_{2}\left(  \beta,E_{\mathbf{k}}%
,\zeta\right)  \right]  ,\label{qq}\\
R  &  =\int\frac{d\mathbf{k}}{\left(  2\pi\right)  ^{3}}\left[  \frac
{f_{1}\left(  \beta,E_{\mathbf{k}},\zeta\right)  +\left(  E_{\mathbf{k}}%
^{2}-3\xi_{\mathbf{k}}^{2}\right)  f_{2}\left(  \beta,E_{\mathbf{k}}%
,\zeta\right)  }{3w}\right. \nonumber\\
&  \left.  +\frac{4\left(  \xi_{\mathbf{k}}^{2}-2E_{\mathbf{k}}^{2}\right)
}{3}f_{3}\left(  \beta,E_{\mathbf{k}},\zeta\right)  +2E_{\mathbf{k}}^{2}%
wf_{4}\left(  \beta,E_{\mathbf{k}},\zeta\right)  \right]  . \label{rr}%
\end{align}
The functions $f_{s}\left(  \beta,\varepsilon,\zeta\right)  $ are the sums
over fermion Matsubara frequencies:%
\begin{equation}
f_{s}\left(  \beta,E,\zeta\right)  =\frac{1}{\beta}\sum_{n=-\infty}^{\infty
}\frac{1}{\left(  \omega_{n}+i\zeta\right)  ^{2}+\varepsilon^{2}}. \label{MS}%
\end{equation}
In Ref. \cite{KTLD2015}, they are analytically determined using the recurrence
relations:%
\begin{align}
f_{1}\left(  \beta,\varepsilon,\zeta\right)   &  =\frac{1}{2\varepsilon}%
\frac{\sinh(\beta\varepsilon)}{\cosh(\beta\varepsilon)+\cosh(\beta\zeta
)},\label{msum}\\
f_{s+1}\left(  \beta,\varepsilon,\zeta\right)   &  =-\frac{1}{2s\varepsilon
}\frac{\partial f_{s}\left(  \beta,\varepsilon,\zeta\right)  }{\partial
\varepsilon}.
\end{align}

The range of validity of EFT has been carefully discussed in our preceding
work \cite{LAKT2015}. The incorporation of rotation in the gradient expansion
is described in Ref. \cite{RotVort1}. The gradient expansion for the two-band
fermion system is performed for each band-component exactly in the same way as
for the one-band system.

\newpage%

\newpage%

\newpage%

\newpage%

\newpage%

\newpage%

\newpage%

\begin{thebibliography}{99}                                                                                               %


\bibitem {Bloch2008}I. Bloch, J. Dalibard, and W. Zwerger, Rev. Mod. Phys.
\textbf{80}, 885 (2008).

\bibitem {zw1}M. W. Zwierlein, J. R. Abo-Shaeer, A. Schirotzek, C. H. Schunck,
and W. Ketterle, Nature (London) \textbf{435}, 1047 (2005).

\bibitem {zw2}M. W. Zwierlein, A. Schirotzek, C. H. Schunck, and W. Ketterle,
Science \textbf{311}, 492 (2006).

\bibitem {zw3}M. W. Zwierlein, C. H. Schunck, A. Schirotzek, and W. Ketterle,
Nature (London) \textbf{442}, 54 (2006).

\bibitem {zw4}C. H. Schunck, M. W. Zwierlein, A. Schirotzek, and W. Ketterle,
Phys. Rev. Lett. \textbf{98}, 050404 (2007).

\bibitem {Clancy}B. Clancy, L. Luo, and J. E. Thomas,Phys. Rev. Lett.
\textbf{99}, 140401 (2007)

\bibitem {Riedl}S. Riedl, E. R. S\'{a}nchez Guajardo, C. Kohstall, J. Hecker
Denschlag, and R. Grimm, New J. Phys. \textbf{13}, 035003 (2011).

\bibitem {Bruun}G. M. Bruun and L. Viverit, Phys. Rev. A \textbf{64}, 063606 (2001).

\bibitem {Cozzini}M. Cozzini and S. Stringari, Phys. Rev. Lett. \textbf{91},
070401 (2003).

\bibitem {Urban}M. Urban, Phys. Rev. A \textbf{71}, 033611 (2005).

\bibitem {Warringa}H. J. Warringa and A. Sedrakian, Phys. Rev. A \textbf{84},
023609 (2011).

\bibitem {Warringa2}H. J. Warringa, Phys. Rev. A \textbf{86}, 043615 (2012).

\bibitem {Wei2012}R. Wei and E. J. Mueller, Phys. Rev. Lett. \textbf{108},
245301 (2012).

\bibitem {Simonucci2015}S. Simonucci, P. Pieri, and G. C. Strinati, Nature
Physics \textbf{11}, 941 (2015).

\bibitem {RotVort1}S. N. Klimin, J. Tempere, N. Verhelst, and M. V.
Milo\v{s}evi\'{c}, Phys. Rev. A \textbf{94}, 023620 (2016).

\bibitem {Iskin2006}M. Iskin and C. A. R. S\'{a} de Melo, Phys. Rev. B
\textbf{74}, 144517 (2006).

\bibitem {Iskin2005}M. Iskin and C. A. R. S\'{a} de Melo, Phys. Rev. B
\textbf{72}, 024512 (2005).

\bibitem {He2015}L. He, H. Hu, and X. J. Liu, Phys. Rev. A \textbf{91}, 023622 (2015).

\bibitem {Pagano2015}G. Pagano, M. Mancini, G. Cappellini, L. Livi, C. Sias,
J. Catani, M. Inguscio, and L. Fallani, Phys.Rev. Lett. \textbf{115}, 265301 (2015).

\bibitem {Hofer2015}M. H\"{o}fer, L. Riegger, F. Scazza, C. Hofrichter, D. R.
Fernandes, M. M. Parish, J. Levinsen, I. Bloch, and S. F\"{o}lling, Phys. Rev.
Lett. \textbf{115}, 265302 (2015).

\bibitem {Zhang2015}R. Zhang, Y. Cheng, H. Zhai, and P. Zhang, Phys. Rev.
Lett. \textbf{115}, 135301 (2015).

\bibitem {Xu2016}J. Xu, R. Zhang, Y. Cheng, P. Zhang, R. Qi, and H. Zhai,
Phys. Rev. A \textbf{94}, 033609 (2016).

\bibitem {Kasahara}S. Kasahara, T. Watashige, T. Hanaguri, Y. Kohsaka, T.
Yamashita, Y. Shimoyama, Y. Mizukami, R. Endo, H. Ikeda, K. Aoyama, T.
Terashima, S. Uji, T. Wolf, H. v. L\"{o}hneysen, T. Shibauchi, and Y. Matsuda,
Proc. Natl. Acad. Sci. USA \textbf{111}, 16309 (2014).

\bibitem {Sun}Y. Sun, W. Zhang, Y. Xing, F. Li, Y. Zhao, Z. Xia, L. Wang, X.
Ma, Q.-K. Xue, and J. Wang, Sci. Rep. \textbf{4}, 6040 (2014).

\bibitem {M2012}J. C. Pi\~{n}a, C. C. de Souza Silva, and M. V.
Milo\v{s}evi\'{c}, Phys. Rev. B \textbf{86}, 024512 (2012).

\bibitem {Chibotaru}L. F. Chibotaru and V. H. Dao, Phys. Rev. B \textbf{81},
020502(R) (2010).

\bibitem {Babaev2002}E. Babaev, Phys. Rev. Lett. \textbf{89}, 067001 (2002).

\bibitem {Simonucci2014}S. Simonucci and G. C. Strinati, Phys. Rev. B
\textbf{89}, 054511 (2014).

\bibitem {Sensarma}R. Sensarma, M. Randeria, and T. L. Ho, Phys. Rev. Lett.
\textbf{96}, 090403 (2006).

\bibitem {KTD}S. N. Klimin, J. Tempere, and J. T. Devreese, Physica C
\textbf{503}, 136 (2014).

\bibitem {KTLD2015}S. N. Klimin, J. Tempere, G. Lombardi, and J. T. Devreese,
Eur. Phys. Journal B \textbf{88}, 122 (2015).

\bibitem {KTD2014}S. N. Klimin, J. Tempere, and J. T. Devreese, Phys. Rev. A
\textbf{90}, 053613 (2014).

\bibitem {LAKT2015}G. Lombardi, W. Van Alphen, S. N. Klimin, and J. Tempere,
Phys. Rev. A \textbf{93}, 013614 (2016).

\bibitem {deMelo1993}C. A. R. S\'{a} de Melo, M. Randeria, and J.R.
Engelbrecht, Phys. Rev. Lett. \textbf{71}, 3202 (1993).

\bibitem {Diener2008}R. B. Diener, R. Sensarma, and M. Randeria, Phys. Rev. A
\textbf{77}, 023626 (2008).

\bibitem {Marini1998}M. Marini, F. Pistolesi, and G. C. Strinati, Eur. Phys.
J. B \textbf{1}, 151 (1998).

\bibitem {Schakel}A. M. J. Schakel, Ann. Phys. \textbf{326}, 193 (2011).

\bibitem {Boettcher}I. Boettcher, J. M. Pawlowski, and S. Diehl, Nucl. Phys.
B, Proc. Suppl. \textbf{228}, 63 (2012).

\bibitem {Diehl}S. Diehl and C. Wetterich, Nucl. Phys. B \textbf{770}, 206 (2007).

\bibitem {Babaev2012-2}M. Silaev and E. Babaev, Phys. Rev. B \textbf{85},
134514 (2012).

\bibitem {KomendovaPRL108}L. Komendov\'{a}, Y. Chen, A.A. Shanenko, M.V.
Milo\v{s}evi\'{c}, and F. M. Peeters, Phys.\ Rev. Lett. \textbf{108}, 207002 (2012).

\bibitem {OurPRA2008}J. Tempere, S. N. Klimin, and J. T. Devreese, Phys. Rev.
A \textbf{78}, 023626 (2008).

\bibitem {Iskin2016}M. Iskin, Phys. Rev. A \textbf{94}, 011604(R) (2016).

\bibitem {Iskin2017}M. Iskin, Phys. Rev. A \textbf{95}, 013618 (2017).

\bibitem {He}L. He, J. Wang, S. Peng, X. J. Liu, and H. Hu, Phys. Rev. A
\textbf{94}, 043624 (2016).
\end{thebibliography}
\end{document}